\documentclass[12pt]{article}
\usepackage{times,amsfonts,amssymb,amsmath,bbm,stmaryrd,amscd,latexsym}
\usepackage[dvips]{graphicx}
\def\e{\epsilon}

\def\g{\gamma}

\def\k{\kappa}
\def\la{\lambda}

\def\om{\omega}
\def\th{\theta}

\def\sfrac#1#2{{\textstyle\frac{#1}{#2}}}

\def\sint{{\textstyle\int\!\mathrm{d}^dx\;}}

\def\im{\mathrm{i}}
\def\eu{\mathrm{e}}
\def\diff{\mathrm{d}}
\def\tr{\mathrm{tr}}
\def\pa{\partial}

\def\>{\rangle}
\def\<{\langle}
\def\+{\dagger}
\def\={\ =\ }
\newcommand{\R}{\mathbb R}

\newcommand{\unity}{{\mathbbm{1}}}
\newcommand{\sTr}{\mathrm{sTr}}
\newcommand{\sDet}{\mathrm{sDet}}
\newcommand{\be}{\begin{equation}}
\newcommand{\ee}{\end{equation}}
\newcommand{\bea}{\begin{eqnarray}}
\newcommand{\eea}{\end{eqnarray}}
\newcommand{\bal}{\begin{aligned}}
\newcommand{\eal}{\end{aligned}}

\newcommand{\und}{\qquad{\textrm{and}}\qquad}
\newcommand{\cb}{}
\newcommand{\cv}{}
\newcommand{\cre}{}

\begin{document}
\begin{titlepage}
\begin{flushright}      %%   to be removed for
ITP--UH--22/13\\        %%   the proceedings
\end{flushright}        %%

\vskip 2.0cm

\begin{center}
{\Large\bf On the Gribov problem in Yang--Mills theory}

\vskip 1.5cm

{\Large 
Olaf Lechtenfeld%$\,{}^{\ddag}$
%$^{}\footnote{E-mail: lechtenf@itp.uni-hannover.de}$
}

\vspace{0.5cm}

\noindent %${}^{\ddag}$
{\it
Institut f\"ur Theoretische Physik and Riemann Center for Geometry and Physics, \\
Leibniz Universit\"at Hannover,
Appelstrasse 2, 30167 Hannover, Germany}

\vspace{0.3cm}

URL: www.itp.uni-hannover.de/\~{}lechtenf

\vspace{1.5cm}

\begin{abstract}
\noindent
I briefly review the Gribov ambiguity of Yang--Mills theory,
some of its features and attempts to control it, in particular
the Gribov--Zwanziger proposal to restrict the functional integration
in the Landau gauge to the Gribov region. This proposal is extended
to an arbitrary gauge in such a way as to guarantee BRST invariance.
The key insight is that any gauge change in the generating functional
can be effected by a suitable field-dependent BRST transformation.
I derive a simple analytic formula for the Jacobian of such a
transformation, which yields an explicit recipe for the required
transformation-parameter functional and allows for the computation
of the Gribov horizon functional in any gauge, as I illustrate for
the class of $R_\xi$ gauges.
\end{abstract}

\end{center}

\vfill

Talk presented at SQS-13 during 29 July -- 03 August, 2013, at JINR, Dubna, Russia
% to be removed for the proceedings

\end{titlepage}

%%%%%%%%%%%%%%%%%%%%%%%%%%%%%%%%%%%%%%%%%%%%%%%%%%%%%%%%%%%%%%%%%%
%\pacs{04.60.Gw,11.30.Pb}     % max. 2 codes of Physics and Astronomy Classification Scheme
%\keywords{Yang-Mills theory, Gribov problem, Gribov--Zwanziger theory, field-dependent BRST transformation} 
%%%%%%%%%%%%%%%%%%%%%%%%%%%%%%%%%%%%%%%%%%%%%%%%%%%%%%%%%%%%%%%%%%

\newpage

\section{What is the Gribov ambiguity?}

Gauge theories are systems with redundant field variables. The simplest prototype is
electrodynamics in $d$ spacetime dimensions, described by gauge potentials 
$\cb A(x)=A_\mu(x)\diff x^\mu$ with $\mu=0,1,\ldots,d{-}1$, subject to gauge transformations
\be
A \mapsto {}^U\!A=U(\diff+A)U^\+ \qquad\textrm{with}\qquad U(x)=\eu^{\im\xi(x)}\ .
\ee
The configurations gauge equivalent to a given~$A$ form the gauge orbit
$\cb {\cal O}_A=\{A'|\,\exists U:A'={}^U\!A\}$,
and the physical configuration space $\cal P$ is the space of gauge orbits or,
equivalently, the quotient space $\cb {\cal P} = \{A\}/\{U\}$, which is 
a topologically and geometrically complicated infinite-dimensional orbifold.
The gauge redundancy is already relevant perturbatively, for the kinetic operator 
in the action possesses zero modes,
\be
(S^{(2)})_{\mu\nu}\,\pa^\nu\!f\equiv(\eta_{\mu\nu}\square-\pa_\mu\pa_\nu)\,\pa^\nu\!f=0\ ,
\ee
hence $S^{(2)}$ is not invertible on $\{A\}$.
To proceed computationally, one needs to fix a gauge, which is a prescription of picking 
a representative~$A$ from each gauge orbit, uniquely and completely,
by a (local) `ideal' condition $\chi(A,x)=0$.
This is implemented in the generating-functional path integral $Z(J)$ via the Faddeev-Popov trick:
Insert \ $1=\int\!{\cal D}U\ \delta\bigl(\chi({}^U\!A)\bigr)\,\bigl|\det K(A)\bigr|$ \ with \
$K(A)=\sfrac{\delta(\chi({}^U\!A))}{\delta U}\big|_{\chi=0}$, factor off the gauge-group volume
and obtain
\be \label{Z}
\begin{aligned}
\cb Z(J)&\cb\= \int\!{\cal D}A\ \delta\bigl(\chi(A)\bigr)\bigl|\det K(A)\bigr|\,
\eu^{\frac{\im}{\hbar}[S_0(A)+\int\!JA]}\\
&\cb\= \int\!{\cal D}(A,C,\bar{C},B)\
\eu^{\frac{\im}{\hbar}[S_0(A)+\int\!\bar{C}KC+\int\!\chi(A)B+\int\!JA]}\ .
\end{aligned}
\ee

Let us illustrate potential pitfalls in a two-dimensional toy example~\cite{VZreview},
with `gauge fields' $(x)=(r,\th)\in\R^2$ subject to `gauge transformations' $\th\mapsto\th+\phi$.
The gauge-fixing function $\chi$ might vanish for more than one angle $\th_i=\th_i(r)$, so in general
\be
\int\!\diff\phi\ \delta\bigl(\chi(r,\th{+}\phi)\bigr)=
\sum_i\bigl|\sfrac{\pa\chi}{\pa\th}\bigl(r,\th_i(r)\bigr)\bigr|^{-1}
\qquad\textrm{with}\qquad \chi(r,\th_i(r))=0\ .
\ee
If the `action function' $S=S(r)$ is independent of $\th$ (`gauge invariance'), then
\be
Z \= \int\!\diff^2x\ \eu^{\im S(r)} \=
\underbrace{\smallint\!\diff\phi}_{2\pi}\int\!\diff^2x\
\Bigl[\sum_i\bigl|{\textstyle\frac{\pa\chi}{\pa\th}}\bigl(r,\th_i(r)\bigr)\bigr|^{-1}\Bigr]^{-1}
\underbrace{\delta\bigl(\chi(x)\bigr)}_{\textrm{kills $\int\!\diff\th$}}\,\eu^{\im S(r)}\ ,
\ee
where the number of terms in the sum may vary with~$r$.
This expression raises a few delicate issues: 
First, how does one take into account the possibility of $\chi$ intersecting with part of the gauge orbits
more than once? Second, does the orientation of the intersection play a role? Third, can I really ignore
the sign of $\det K(A)$ in~(\ref{Z})?

An important insight was achieved by Gribov~\cite{Gribov}, who realized that, in Yang-Mills theory, 
where $A=A^aT^a\in su(n)$, the Landau gauge $\chi=\pa{\cdot}A$ 
(like, in fact, any covariant gauge) is not ideal in the above sense, 
because a gauge orbit ${\cal O}_A$ may contain more than one configuration with $\chi=0$:
\be
\exists A' : \quad A'\=U(\diff+A)U^\+ \qquad\textrm{with}\qquad \pa^\mu\!A_\mu=0=\pa^\mu\!A'_\mu\ .
\ee
If $A'$ is infinitesimally close to $A$, one may approximate $U=\eu^\xi=\eu^{\xi^aT^a}\simeq\unity+\xi$,
and the existence of a so-called Gribov copy is equivalent to
\be
0\=\pa^\mu(\pa_\mu+\textrm{ad}A_\mu)\xi\=\pa^\mu\!D_\mu\xi\ .
\ee
This condition on $A$ simply means that the Faddeev-Popov operator
$K(A)=-\pa^\mu\!D_\mu$ (here in Landau gauge) possesses a non-constant zero mode.
When $A$ is `small', i.e.~in perturbation theory, $K(A)\simeq-\pa^\mu\pa_\mu$
has only positive nontrivial eigenvalues, so no Gribov problem occurs. For the
same reason, the problem does not appear in QED altogether. However, when $A$
becomes `large enough', $K(A)$ eventually develops negative eigenvalues.
At some critical `size' of $A$, some eigenvalue of~$K(A)$ crosses zero, meaning
that $\det K(A)$ switches sign and a new Gribov copy will appear.
Gribov~\cite{Gribov} was the first to realize that many popular gauges (like Landau or Coulomb)
yield infinitely many Gribov copies. Since then, such (infinitesimal) Gribov copies 
have been constructed rather explicitly.

It is customary to define the `Gribov region' and `first Gribov horizon' by
\be
\Omega\ :=\ \{ A\,|\ \chi(A)=0\ \ \&\ \ K(A)>0\} \und 
\pa\Omega \ ,
\ee
respectively. Then, infinitesimal Gribov copies sit on either side of the Gribov horizon $\pa\Omega$.
Let me list some important properties of the Gribov region, disregarding for this
purpose {\it global\/} gauge transformations:
\begin{itemize}
\item
An alternative definition is \quad 
$\Omega\=\{\textrm{relative minima of} \quad 
\|A\|^2=\tr\int\!\diff^dx\ A{\cdot}A \quad\textrm{on ${\cal O}_A$}\}$\\
because \quad $0=\delta\|A\|^2\ \Leftrightarrow\ \pa{\cdot}A=0$ \quad
and \quad $0<\delta^2\|A\|^2\ \Leftrightarrow\ -\pa{\cdot}D>0$.
\item
$\|A\|^2$ achieves its {\cre absolute minimum} on each gauge orbit, 
thus each gauge orbit intersects~$\Omega$.
\item
$\Omega$ is convex and bounded in every direction, because \quad
$\lim_{\la\to\infty}K(\la A)=\la\,\pa^\mu\textrm{ad}A_\mu$ \quad
is traceless. Hence, there exists a negative eigenvalue, so $\la A\notin\Omega$.
\item
The Gribov region still contains Gribov copies! The reason is that
$\|A\|^2$ on ${\cal O}_A$ develops a saddle point at $\pa\Omega$.
Therefore, $\|A\|^2$ can be lowered inside $\Omega$. Clearly,
a gauge orbit ${\cal O}_A$ can feature more than one relative minimum of $\|A\|^2$.
\end{itemize}
In view of the above complications, a more strict notion is useful, and one defines
the `fundamental modular region' (FMR) as \quad
\be
\Lambda\ :=\ \{\textrm{absolute minimum of} \ \|A\|^2 \ \textrm{on ${\cal O}_A$}\}\ .
\ee
It follows that $\Lambda\subset\Omega$, \ ${\cal O}\cap\Lambda\neq0$, \
$\Lambda$ is convex and bounded in all directions, and $\pa\Lambda\cap\pa\Omega\neq0$.\\
One should note, however, that {\it degenerate\/} absolute minima of $\|A\|^2$ live
on $\pa\Lambda$, and so the boundary $\pa\Lambda$ still holds Gribov copies!
Most of the historical material presented in this and the following section are taken
from the review by Vandersickel and Zwanziger~\cite{VZreview}.

\section{Proposals for a remedy}

Over the years, a fair number of proposals have been put forward to control
the Gribov copies. The ubiquity of the feature is epitomized by Singer's theorem~\cite{Singer},
which states that only singular, i.e.~non-continuous, gauges can be free of Gribov copies.

An obvious reaction is to select such a singular gauge, 
for example a space-like planar or a hyperaxial one. 
Yet, these are very cumbersome, and the computational price may be too high.

It has been suggested to take (\ref{Z}) literally and to lift the {\it absolute value\/}
of $\det K$ into the action, but it remains unclear whether this procedure properly
accounts for the number of Gribov copies.
The opposite recipe maintains $\det K$ {\sl without absolute value\/} and integrates 
over all Gribov copies, hoping that alternating orientations of the intersection
of ${\cal O}_A$ with the gauge condition will lead to a cancellation between most copies.

Other ideas invoke stochastic quantization, which introduces a `gauge-fixing force'
tangential to the gauge orbits, or simply a {\it restriction of the functional integration\/} 
to the Gribov region~$\Omega$ or the FMR~$\Lambda$.
The latter is connected with a hope for a confinement mechanism:
Since $\Omega$ is compact, quantization might give rise to a mass gap.

It is believed that the boundary of $\Omega$ carries a lot of weight, namely that
the path integral is dominated by {\it degenerate orbits\/}
\be
{\cal O}_{\bar A} \quad \textrm{for}\quad \bar{A} \quad\textrm{with}\quad 
D(\bar{A})\xi \= 0 \qquad\Longrightarrow\qquad K(\bar{A})\=0
\qquad\Longrightarrow\qquad \bar{A}\in\partial\Omega\ ,
\ee
so that dim ${\cal O}_{\bar A}$ is smaller than the generic dim ${\cal O}_A$ 
by the number of solutions~$\xi$.

Denoting by $\k(A)$ the lowest eigenvalue of $K(A)$ and cutting off the
functional integration at~$\partial\Omega$, 
in a saddle-point approximation one obtains~\cite{VZreview}
\be
\begin{aligned}
Z(J) &\= \int\!{\cal D}(A,C,\bar{C},B)\ \th[\k(A)]\
\eu^{\frac{\im}{\hbar}[S_0(A)+\int\!\bar{C}KC+\int\!\chi(A)B+\int\!JA]} \\
&\= \sfrac{1}{2\pi\im}\smallint_{-\infty}^{\infty}\sfrac{\diff\om}{\om{-}\im\e}
\int\!{\cal D}(A,C,\bar{C},B)\ \eu^{\im\om\k(A)}\
\eu^{\frac{\im}{\hbar}[S_0(A)+\int\!\bar{C}KC+\int\!\chi(A)B+\int\!JA]} \\
&\!\!\!\!\buildrel{\textrm{\tiny saddle point}}\over{\simeq}
\int\!{\cal D}(A,C,\bar{C},B)\
\eu^{\frac{\im}{\hbar}[S_0(A)+\int\!\bar{C}KC+\int\!\chi(A)B+\g^2 H(A,C,\bar{C},B)+\int\!JA]}
\end{aligned}
\ee
with a `horizon functional' $H$. In the Landau gauge, $\chi=\partial{\cdot}A$, 
the latter depends on $A$ only and has been computed to be~\cite{VZreview}
\be \label{HLandau}
H(A)\=\int\!\!\diff^dx\!\int\!\!\diff^dy\
f^{abc}A^b_{\mu}(x)(K^{-1})^{ad}(x{-}y)f^{dec}A^{e\mu}(y)
\ -\ \int\!\!\diff^dx\ d(n^2{-}1)\ ,
\ee
where $f^{abc}$ denote the gauge-group structure constants.
The `Gribov parameter' $\g$ is to be determined self-consistently
via the `gap equation'
\be \label{gap}
\sfrac{\pa\ln Z(0)}{\pa\g}(\g)=0 \qquad{\Leftrightarrow}\qquad
\bigl\< H(A) \bigr\>_\g = 0 \qquad \textrm{`horizon condition'}\ .
\ee

Three remarks are in order:
Firstly, the integration measure above peaks around $\pa\Omega$, supporting the
`degenerate-orbit dominance' hypothesis.
Secondly, $\g\sim\exp\{-1/g^2\}$ \ vanishes perturbatively, 
so its effect is only seen in the infrared.
Thirdly, one finds that the gluon propagator behaves as $\frac{k^2}{k^4+2{\cv\g^2}g^2N}$
while the ghost propagator gets enhanced like~$\frac{1}{k^4}$,
consistent with the mass-gap picture.

\section{Yang--Mills theory in Faddeev-Popov quantization}

Almost all considerations regarding the Gribov problem have been made in the Landau gauge.
However, for any proposal of overcoming the problem in a specific gauge, there arises
the crucial issue of gauge invariance. It is therefore necessary to probe such proposals
for nearby (or even distant) other gauges~\cite{LLR}. The proper tool for achieving this is 
a gauge-changing procedure for the generating functional, preferably in the BRST formulation.

I begin by reminding the audience of the salient features of the Faddeev-Popov quantization
of SU($n$) Yang-Mills theory in $\R^{1,d-1}$. Its classical action reads
\be
S_0(A) \= -\sfrac14\int\!\diff^d x\ F_{\mu\nu}^{a}F^{\mu\nu{}a}
\qquad\textrm{with}\qquad
F^a_{\mu\nu}\=\partial_{\mu}A^a_{\nu}-\partial_{\nu}A^a_{\mu}+
f^{abc}A^b_{\mu}A^c_{\nu}\ ,
\ee
where $a=1,\ldots,n$ \ and \ $\mu=0,1,\ldots,d{-}1$.
$S_0$ is invariant under gauge transformations 
\be
\delta A^a_{\mu}\=D^{ab}_{\mu}\xi^b \qquad\textrm{with}\qquad
D^{ab}_{\mu}\=\delta^{ab}\partial_{\mu}+f^{acb}A^c_{\mu}
\und \xi^b=\xi^b(x)\ .
\ee
The BRST formulation of the quantum theory extends the field content and the action to
\be
\bigl\{\phi^A\bigr\}\=\bigl\{A^a_{\mu}(x), B^a(x), C^a(x), {\bar C}^a(x)\bigr\}\ ,
\ee
\be
S(\phi)\= S_0(A)\ +\ \sint{\bar C}^a K^{ab}(A)\,C^b\ +\ \sint\chi^a(A)\,B^a\ ,
\ee
with a choice $\cb\chi^a$ for a gauge-fixing function and 
with the ensuing Faddeev-Popov operator 
\be
K^{ab}(A)\=\frac{\partial\chi^a(A)}{\partial A^c_{\mu}}D^{cb}_{\mu}\ .
\ee
In the Landau gauge, $\chi^a=\pa{\cdot}A^a$, one has 
$K^{ab}(A)=\delta^{ab}\partial^{\mu}\partial_{\mu}+ f^{acb}A^c_{\mu}\partial^{\mu}$.

The extended action $S$ is invariant under (even) BRST transformations
\be
\delta_{\lambda} A_{\mu}^{a} = D^{ab}_{\mu}C^b\lambda\ ,\quad
\delta_{\lambda} \bar{C}{}^a = B^a\lambda\ ,\quad
\delta_{\lambda} B^a = 0\ ,\quad
\delta_{\lambda} C^a = \sfrac12 f^{abc}C^bC^c\lambda\ ,
\ee
where $\lambda$ is an odd constant. It is convenient to introduce
the (odd) Slavnov variation $sX$ of any functional~$X$ by writing
\be 
\delta_{\lambda}X(\phi)\=\big(sX(\phi)\big)\,\lambda
\qquad\textrm{so that}\qquad
sX(\phi)\=\frac{\delta X(\phi)}{\delta\phi^A}R^A(\phi)\ ,
\ee
with the combined short-hand notation
\be 
\bigl\{R^A(\phi)\bigr\}\=
\bigl\{D^{ab}_{\mu}C^b(x)\;,\; 0\;,\; \sfrac12 f^{abc}C^bC^c(x)\;,\; B^a(x)\bigr\}
\ee
and DeWitt's extension~\cite{DeWitt} of Einstein's summation convention 
(sum over $A$ includes integration over $x$).
The nilpotency of the Slavnov variation, $s^2=0$, implies that 
\be 
0 \= s R^A(\phi) \= \smash{\frac{\delta R^A(\phi)}{\delta\phi^B}}R^B(\phi)
\ \equiv\ R^A_{\ ,B}\,R^B\ .
\ee
It is very useful to define the extended (odd) gauge-fixing functional
\be
\psi(\phi)\=\sint{\bar C}^a\chi^a(A)\ ,
\ee
in terms of which the extended action can be made manifestly BRST invariant:
\be \label{eYM}
S(\phi)\= S_0(A)\ +\ \sint{\bar C}^a K^{ab}(A)\,C^b\ +\ \sint\chi^a(A)\,B^a
\=S_0(A)\ +\ s\psi(\phi)\ =:\ S_\psi(\phi)\ ,
\ee
thus obviously $sS_\psi(\phi)=0$.

\section{Field-dependent BRST transformations}

The main point of this talk, based on~\cite{LL1,LL2}, 
identifies a gauge change with a field-dependent BRST transformation. 
Let me therefore generalize the odd constant~$\lambda$ to a 
{\it field-dependent\/} (but still $x$ independent) 
BRST-parameter {\it functional\/} $\Lambda(\phi)$. 
The corresponding transformation then reads
\be 
\delta_\Lambda X(\phi) \= \bigl( sX(\phi)\bigr) \Lambda(\phi)
\= X_{,A} R^A \Lambda(\phi) \qquad\textrm{with}\qquad \Lambda^2(\phi)=0\ .
\ee
On the fields $\phi^A$ themselves, this amounts to a (nonlocal) 
change of field variables $\phi\to\varphi$,
\be
\varphi^A\=\varphi^A(\phi)
\=\phi^A+\delta_\Lambda\phi^A
\=\phi^A+(s\phi^A)\Lambda(\phi)
\=\phi^A+R^A(\phi)\Lambda(\phi)\ ,
\ee
with a Jacobian supermatrix
\be
\begin{aligned}
M^A_{\ \ B}(\phi)\=\frac{\delta \varphi^A(\phi)}{\delta\phi^B}
&\=\delta^A_{\ \ B}\ +\
\frac{\delta R^A(\phi)}{\delta\phi^B}\Lambda(\phi)(-1)^{\varepsilon_B}\ +\
R^A(\phi)\frac{\delta \Lambda(\phi)}{\delta\phi^B} \\[4pt]
&\ \equiv\ \delta^A_{\ \ B}\ +\
R^A_{\ ,B}\Lambda(-1)^{\varepsilon_B}\ +\ R^A\Lambda_{,B}\ .
\end{aligned}
\ee
Surprisingly, its superdeterminant can be computed {\it exactly\/}:
\be
\begin{aligned}
\sTr\ln M(\phi) &\=
-\sum_{n=1}^\infty\frac{(-1)^n}{n}\ \sTr\bigl(
R^A_{\ \ ,B}\Lambda(-1)^{\varepsilon_B}+R^A\Lambda_{,B}\bigr)^n \\
&\=-\sum_{n=1}^\infty\frac{(-1)^n}{n}\ \sTr\bigl( R^A\Lambda_{,B}\bigr)^n\\
&\=+\sum_{n=1}^\infty\frac{(-1)^n}{n}\bigl(\Lambda_{,A} R^A\bigr)^n \\
&\= \sum_{n=1}^\infty\frac{(-1)^n}{n}(s\Lambda)^n\\
&\=-\ln\bigl(1+s\Lambda(\phi)\bigr)\ ,
\end{aligned}
\ee
hence
\be
\sDet M(\phi)\=\bigl[1+s\Lambda(\phi)\bigr]^{-1}\ .
\quad\qquad\qquad\qquad\qquad\qquad{}
\ee

Performing such a variable change in a functional integral, one obtains
\be
\begin{aligned}
{\cal I}
&\=\int\!{\cal D}\varphi\;\exp\Big\{\sfrac{\im}{\hbar}W(\varphi)\Big\}\\
&\=\int\!{\cal D}\phi\ \;\sDet M(\phi)\;
\exp\Big\{\sfrac{\im}{\hbar}W(\varphi(\phi))\Big\}\\
&\=\int\!{\cal D}\phi\;\exp\Big\{\sfrac{\im}{\hbar}\big[W(\varphi(\phi))-\im\hbar\
\sTr\ln M(\phi)\big]\Big\}\\
&\=\int\!{\cal D}\phi\;\exp\Big\{\sfrac{\im}{\hbar}
\big[W(\phi)+\bigl(sW(\phi)\bigr)\Lambda(\phi)+
\im\hbar\,\ln\bigl(1+s\Lambda(\phi)\bigr)\big]\Big\}\ ,
\end{aligned}
\ee
hence the functional $W$ is shifted by a classical and a quantum piece.
It is important to realize that these transformations are {\it not\/} nilpotent,
since
\be
\delta_\Lambda^2 X(\phi)
\=\delta_\Lambda\bigl[ \bigl(sX(\phi)\bigr)\Lambda(\phi)\bigr]
\=\bigl(sX(\phi)\bigr)\bigl(s\Lambda(\phi)\bigr)\Lambda(\phi)
\ee
vanishes only if
\be
0 \= s\Lambda(\phi) \= \Lambda_{,A}(\phi)R^A(\phi)\ ,
\ee
which of course includes the trivial case of \ $\Lambda(\phi)=\lambda=\textrm{constant}$.

\section{Relating different gauges}

I will now be more specific and take for $W$ the extended Yang--Mills action~(\ref{eYM}),
i.e.~investigate the form change of the Yang--Mills vacuum functional in a gauge $\psi$ under 
the change of field variables induced by a field-dependent BRST transformation:
\be
\begin{aligned}
Z_\psi(0)
&\=\int\!{\cal D}\varphi\; \exp\Big\{\sfrac{\im}{\hbar}S_\psi(\varphi)\Big\}\\
&\!\!\!\buildrel{sS_\psi=0}\over{=}\!\!\int\!{\cal D}\phi\; \exp\Big\{\sfrac{\im}{\hbar}
\big[S_\psi(\phi)+\im\hbar\,\ln\big(1+s\Lambda(\phi)\big)\big]\Big\}\\
&\=\int\!{\cal D}\phi\;\exp\Big\{
\sfrac{\im}{\hbar}\big[S_0(A)+s\,\psi(\phi)+s\,\delta\psi(\phi)\big]\Big\}
\= Z_{\psi+\delta\psi}(0)
\end{aligned}
\ee
where the last term in the exponent is BRST exact,
\be
\im\hbar\,\ln\big(1+s\Lambda(\phi)\big)\=s\,\delta\psi(\phi) \qquad\textrm{with}\qquad
\delta\psi(\phi)\=
\im\hbar\,\Lambda(\phi)\bigl(s\Lambda(\phi)\bigr)^{-1}\ln\big(1+s\Lambda(\phi)\big)\ .
\ee
This shows that a field-dependent BRST transformation with $\Lambda$ 
effects a shift of the extended gauge-fixing functional by some 
$\delta\psi\simeq\im\hbar\Lambda+O(\Lambda^2)$.

It is illuminating to reverse the dependence and determine which BRST-parameter functional
$\Lambda$ has to be chosen in order to achieve a given gauge change $\delta\psi$.
The inversion 
\be 
s\Lambda(\phi) \= \exp\big\{\sfrac{1}{\im\hbar}s\,\delta\psi\bigr\}-1 
\ee
is solved by (see also~\cite{JM})
\be
\begin{aligned}
\Lambda(\phi) &\= \delta\psi\,(s\,\delta\psi)^{-1}
\bigl(\exp\big\{\sfrac{1}{\im\hbar}s\,\delta\psi\bigr\}-1\bigr) \\
&\= \sfrac{1}{\im\hbar}\,\delta\psi\!\sum_{n=0}^\infty \sfrac{1}{(n+1)!}
\bigl(\sfrac{s\delta\psi}{\im\hbar}\bigr)^n
\ \simeq\ \sfrac{\delta\psi}{\im\hbar}+
\sfrac{\delta\psi\,s\delta\psi}{2(\im\hbar)^2}+\ldots\ .
\end{aligned}
\ee

A prime example is the class of $R_\xi$ gauges, defined by
\be \label{rxi}
\psi_\xi(\phi)\=\sint{\bar C}^a\bigl(\partial^{\mu}A^a_{\mu}+\sfrac{\xi}{2}B^a\bigr)\ .
\ee
To move from $R_\xi$ to $R_{\xi+\delta\xi}$ gauge needs
\be
\delta\psi=\sfrac12\delta\xi\,\sint{\bar C}^a\!B^a \qquad{\Longrightarrow\qquad}
s\,\delta\psi = \sfrac12\delta\xi\,B^2 \qquad\textrm{with}\quad B^2=\sint B^a\!B^a\ ,
\ee
and the corresponding field-dependent BRST-parameter functional reads
\be
\begin{aligned}
\Lambda(\phi) &\= (B^2)^{-1}
\bigl( \exp\bigl\{\sfrac{\delta\xi}{2\im\hbar}B^2\bigr\}-1 \bigr)
\sint {\bar C}^a\!B^a \\
&\=\sfrac{\delta\xi}{2\im\hbar}\,\Bigl\{
1+\sfrac{1}{2!}\sfrac{\delta\xi}{2\im\hbar}B^2+
\sfrac{1}{3!}\bigl(\sfrac{\delta\xi}{2\im\hbar}B^2\bigr)^2+
\sfrac{1}{4!}\bigl(\sfrac{\delta\xi}{2\im\hbar}B^2\bigr)^3+\ldots \Bigr\}
\sint {\bar C}^a\!B^a
\end{aligned}
\ee
in a (nonlocal) power series expansion in $\delta\xi$ (and $B^2/\hbar$).

\section{Gribov horizon beyond the Landau gauge}

The nonlocal Gribov-Zwanziger model~\cite{VZreview} is defined by adding 
the (non-local) Gribov horizon functional to the extended action~(\ref{eYM}). 
In the Landau gauge, its (non-local) action then reads
\be \label{SH}
S_H(\phi)\=S_\psi(\phi)\ +\ {\cv\g^2} H_0(A)
\=S_0(A)\ +\ s\psi_0(\phi)\ +\ {\cv\g^2} H_0(A)\ ,
\ee
with
\be
\begin{aligned} 
\psi_0(\phi)&\=\sint{\bar C}^a\,\partial{\cdot}A^a \und \\[4pt]
H_0(A)&\=\sint\!{\textstyle\int\!\!\diff^dy}\
f^{abc}A^b_{\mu}(K^{-1})^{ad}f^{dec}A^{e\mu}\ -\ \sint\ d(n^2{-}1)\ ,
\end{aligned}
\ee
and the Gribov parameter $\g$ is determined self-consistently by the gap equation~(\ref{gap}).

As it stands in~(\ref{SH}), the Gribov-Zwanziger action $S_H$ is {\it not\/} 
BRST invariant, because $H$ is defined in the Landau gauge, 
and its Slavnov variation does not vanish,
\be
sH_0\=\sint\!{\textstyle\int\!\!\diff^dy}\ 
f^{abc}f^{cde}\bigl[2D^{bq}_{\mu}C^q(K^{-1})^{ad}-
f^{mpn}{\textstyle\int}A^b_{\mu}(K^{-1})^{am}K^{pq}C^q(K^{-1})^{nd}\bigr]A^{e\mu}\ \neq\ 0\ .
\ee
Does this imply that the effective quantum action and thus the physical S-matrix
are gauge dependent even on-shell?
No, it only means that away from the Landau gauge one cannot use the same horizon functional,
but should properly modify it such as to account for the gauge change.
This modification can now be directly domputed using the tool developed in the previous section.

For the example of the $R_\xi$ gauges~(\ref{rxi}), 
let me move from $\xi{=}0$ to some finite value of~$\xi$.
As I demonstrated, this can be done by a field-dependent BRST transformation,
\be
\phi^A\quad \longmapsto \quad
\phi^A+\,(s\phi^A)\Lambda_\xi(\phi) \qquad\textrm{with}\qquad
\Lambda_\xi(\phi)\=(B^2)^{-1}
\big(\exp\big\{\sfrac{\xi\,B^2}{2\im\hbar}\big\}-1\big) \sint {\bar C}^a\!B^a
\ee
leading to
\be
\begin{aligned}
Z_{\xi=0}(0)&\=\int\!{\cal D}\phi\ \exp\Big\{\frac{\im}{\hbar}\big(
S_0(A)+ s\psi_0(\phi)+\g^2 H_0(A)\big)\Big\} \\
&\=\int\!{\cal D} \phi\ \exp\Big\{\frac{\im}{\hbar}\big(
S_0(A)+s\psi_\xi(\phi)+\g^2 H_\xi(\phi)\big)\Big\}
\= Z_\xi(0)\ ,
\end{aligned}
\ee
where I read off
\be
\psi_\xi(\phi)\=\sint{\bar C}^a\big(\partial^{\mu}\!A^a_{\mu}+\sfrac{\xi}{2}B^a\big)
\qquad\textrm{and}\qquad
H_\xi(\phi)\=H_0(A)+\big(sH_0(A,C)\big)\Lambda_\xi(\phi)\ .
\ee
The last equation provides a proposal for the horizon functional in a general $R_\xi$ gauge
in such a way that gauge invariance is restored in the vacuum functional:
\be
\g^2\bigl\<\delta H(\phi)\bigr\>_\g\ +\ \bigl\<s\,\delta\psi(\phi)\bigr\>_\g\=0
\qquad\textrm{under}\qquad \psi\mapsto\psi+\delta\psi\ .
\ee
To demonstrate the computability, here is its explicit form:
{\small
\begin{eqnarray}
H_\xi(\phi)&=& 
f^{abc}{\textstyle\int}\!{\textstyle\int}A^b_{\mu}(K^{-1}\!)^{ad}f^{dec}A^{e\mu}
\ -\ {\textstyle\int} d(n^2{-}1) \\[4pt] \nonumber
+\ \ f^{abc}\!\!\!\!\!\!\!\!&&\!\!\!\!\!\!\!\! f^{cde} {\textstyle\int}\!{\textstyle\int}
\big[2D^{bq}_{\mu}C^q(K^{-1}\!)^{ad} - f^{mpn}{\textstyle\int}
A^b_{\mu}(K^{-1}\!)^{am}K^{pq}C^q(K^{-1}\!)^{nd}\big] A^{e\mu}
(B^2)^{-1}\big(\eu^{\frac{\xi}{2\im\hbar}B^2}{-}1\big) 
{\textstyle\int}{\bar C}^\ell\! B^\ell\,.
\end{eqnarray}
}

This idea may likewise be applied to the {\it local\/} form of the Gribov-Zwanziger 
model~\cite{VZreview}, which further extends the field space to
\be
\bigl\{\Phi^{\cal A}\bigr\}\=
\bigl\{\phi^A\,,\,\varphi^{ac}_\mu\,,\,{\bar\varphi}^{ac}_\mu\,,\,
\omega^{ac}_\mu\,,\,{\bar\omega}^{ac}_\mu\bigr\}
\ee
and features the following (local) action,
\be
S_{GZ}(\Phi)\=S_0(A)\ +\ s\psi(\phi)\ +\
S_{\g}(A,\varphi,{\bar\varphi},\omega,{\bar\omega})\ ,
\ee
where the (now local) `improvement functional' reads
\be
S_{\g}\=\sint\big[{\bar\varphi}^{ac}_\mu K^{ab}\varphi^{\mu bc}\ -\
{\bar\omega}^{ac}_\mu K^{ab}\omega^{\mu bc}\ +\ 2\im\g f^{abc}A^{b}_{\mu}
\big(\varphi^{\mu ac}+{\bar\varphi}^{\mu ac}\big)\ +\ \g^2\,d(n^2{-}1)\big]\ .
\ee
The new fields form two BRST doublets,
\be
\begin{aligned} 
\delta_{\lambda} \varphi^{ac}_\mu &=\omega^{ac}_\mu\lambda\ ,\qquad\qquad\ \,
\delta_{\lambda} {\bar\varphi}^{ac}_\mu=0\ ,\\ 
\delta_{\lambda} \omega^{ac}_\mu &=0\ ,\qquad\qquad\qquad
\delta_{\lambda} {\bar\omega}^{ac}_\mu =-{\bar\varphi}^{ac}_\mu\lambda\ .
\end{aligned}
\ee
Again, BRST invariance is (softly) broken since
\be
sS_{\g}\=f^{adb}\sint\big[{\bar\varphi}^{ac}_\mu K^{de}C^e\varphi^{\mu bc}+
{\bar\omega}^{ac}_\mu K^{de}C^e\omega^{\mu bc}+
2\im\g\big( D^{de}_{\mu}C^e(\varphi^{\mu ab}{+}{\bar\varphi}^{\mu ab})+
A^d_{\mu}\omega^{\mu ab}\big)\big]\ \neq\ 0\ .
\ee
However, gauge invariance can be restored by defining (for general $R_\xi$ gauges)
\be
S_{\g\xi}(\Phi)\=S_{\g}(A,\varphi,{\bar\varphi},\omega,{\bar\omega})
\ +\ \big(sS_{\g}(A,C,\varphi,{\bar\varphi},\omega,{\bar\omega})\big)\Lambda_\xi(\phi)\ ,
\ee
so that the action changes via the corresponding field-dependent BRST transformation as
\be
S_{GZ}(\Phi)\quad\longmapsto\quad
S_0(A)\ +\ s\psi_\xi(\phi)\ +\ S_{\g\xi}(\Phi)\ ,
\ee
induced by a harmless change of variables in the functional integral.
Once more,
\be
sS_\g(\Phi)\neq0 \qquad\textrm{but}\qquad
\delta S_\g(\Phi)=sS_\g(\Phi)\,\Lambda_\xi(\phi)
\qquad\textrm{balances}\qquad s\,\delta\psi(\phi)\ .
\ee

\section{Outlook}

I have shown how it is possible to move from a reference gauge $\psi_0$ to a 
desired gauge $\psi$ via a field-dependent BRST transformation with parameter functional
\be
\begin{aligned}
\Lambda_\psi(\phi) &\cb\= (\psi{-}\psi_0)\bigl(s(\psi{-}\psi_0)\bigr)^{-1}
\bigl( \exp\bigl\{\sfrac{1}{\im\hbar}s(\psi{-}\psi_0)\bigr\}-1 \bigr)\\
&\= \sfrac{1}{\im\hbar}(\psi{-}\psi_0)
\sum_{n=0}^\infty\sfrac{1}{(n+1)!}\bigl(\sfrac{1}{\im\hbar}s(\psi{-}\psi_0)\bigr)^n\ .
\end{aligned}
\ee
This connection proposes a corresponding change of the horizon functional,
\be
H_\psi(\phi) - H_0(\phi) \= \big(sH_0(\phi)\big)\Lambda_\psi(\phi)\ .
\ee
One may use these results in $R_\xi$ gauges to interpolate between the interpretation-friendly
unitary gauge ($\xi{\to}\infty$) and the renormalization-friendly Landau gauge~($\xi{\to}0$).
It will be also very interesting to go beyond $R_\xi$ gauges and relate, for instance,
the Coulomb gauge to the Landau gauge. 
Finally, I have only discussed the vacuum functional. The analysis should be (and can be)
extended to investigate the gauge variation of Greens functions, VEVs and S-matrix elements 
using
\be
Z_{\psi+\delta\psi}(J) \= Z_\psi(J)\ -\
\sfrac{\im}{\hbar}J_A\,\bigl\<(s\phi^A)\,\Lambda_{\delta\psi}(\phi)\bigr\>_J\ .
\ee

\bigskip

\noindent
{\bf\large Acknowledgments}\\[8pt]
I wish to thank Peter Lavrov for involving me into this project and for the pleasant collaboration.

\end{document}